\documentclass[conference]{IEEEtran}
\IEEEoverridecommandlockouts
\usepackage{cite}
\usepackage{amsmath,amssymb,amsfonts}
\usepackage{algorithmic}
\usepackage{graphicx}
\usepackage{textcomp}
\usepackage{xcolor}
\usepackage{orcidlink}

\def\BibTeX{{\rm B\kern-.05em{\sc i\kern-.025em b}\kern-.08em
    T\kern-.1667em\lower.7ex\hbox{E}\kern-.125emX}}

\usepackage{xspace}
\usepackage{hyperref}
\usepackage{comment}
\usepackage{booktabs}
\usepackage{enumitem}
\usepackage{array}

\newcommand{\name}{\texttt{Uniference}\xspace}

\begin{document}

\title{UNIFERENCE: A Discrete Event Simulation Framework for Developing Distributed AI Models}

\author{\IEEEauthorblockN{Doğaç Eldenk \orcidlink{0009-0004-0628-9156}}
\IEEEauthorblockA{
\textit{Department of Computer Science} \\
\textit{Northwestern University}\\
Evanston, United States \\
dogac@u.northwestern.edu}
\and
\IEEEauthorblockN{Stephen Xia \orcidlink{0000-0001-5713-8885}}
\IEEEauthorblockA{
\textit{Department of Computer Engineering} \\
\textit{Northwestern University}\\
Evanston, United States \\
stephen.xia@northwestern.edu}
}

\maketitle

\begin{abstract}
  Developing and evaluating distributed inference algorithms remains difficult due to the lack of standardized tools for modeling heterogeneous devices and networks. Existing studies often rely on ad-hoc testbeds or proprietary infrastructure, making results hard to reproduce and limiting exploration of hypothetical hardware or network configurations. We present \name, a discrete-event simulation (DES) framework designed for developing, benchmarking, and deploying distributed AI models within a unified environment. \name models device and network behavior through lightweight logical processes that synchronize only on communication primitives, eliminating rollbacks while preserving the causal order. It integrates seamlessly with PyTorch Distributed, enabling the same codebase to transition from simulation to real deployment. Our evaluation demonstrates that \name profiles runtime with up to 98.6\% accuracy compared to real physical deployments across diverse backends and hardware setups. By bridging simulation and deployment, \name provides an accessible, reproducible platform for studying distributed inference algorithms and exploring future system designs, from high-performance clusters to edge-scale devices. The framework is open-sourced at \url{https://github.com/Dogacel/Uniference}.
\end{abstract}


\vspace{-10pt}
\section{Introduction}
\label{introduction}

Recent progress in machine learning has shown that model accuracy continues to improve with scale: larger datasets, parameters, and compute resources consistently yield better performance. However, this scaling law has quickly outpaced hardware capabilities. Modern large language models contain tens of billions of parameters, too large to fit onto a single GPU even in high-end data centers. At the same time, many real-world deployments operate under the opposite constraint. They rely on heterogeneous and resource-limited hardware \cite{hu_distributed_2022}, ranging from older GPUs to embedded and mobile devices. The resulting landscape spans edge to cloud, where computation is fragmented across devices with vastly different compute, memory, and network characteristics. Efficiently serving these models across different environments has therefore become a key systems challenge \cite{bambhaniya_understanding_2025,alizadeh_llm_2024,zheng_review_2025,gill_edge_2025,friha2024llm,li_primacpp_2025}.

While distributed training has received extensive attention \cite{megatron-lm}, distributed inference presents a distinct and relatively under-explored problem. Training is an offline, throughput-oriented process that can tolerate high latency and relies on asynchronous, uniform hardware. On the other hand, inference is online and latency-critical: it must respond in real time to user queries or sensor input \cite{cai_edge-llm_2024}. Moreover, inference workloads often run on different hardware, including high-performance or consumer-grade GPUs, laptops, smartphones, IoT devices, and robots based on connectivity, privacy and cost constraints. In these environments, distributing inference across nearby devices or combining edge and cloud resources greatly improves accuracy and efficiency by utilizing idle or more powerful compute resources \cite{li2023adaptive}.

Numerous techniques have been proposed to partition models and distribute inference. Tensor (TP), pipeline (PP), and expert-parallel (EP) techniques are some of them \cite{hu_pipeline_2021}. Development of such algorithms are complex, and their evaluation remains fragmented and hard to reproduce. Most studies test algorithms on proprietary or limited hardware, using ad-hoc datasets and network setups based on their access to resources \cite{ye_galaxy_2024,li_tpi-llm_2024,qazi_prism_2025,parthasarathy_defer_2022}. Replicating such experiments is often infeasible. Realistic testing requires access to many types of devices, careful control of bandwidth and latency, and extensive environment configuration. These barriers make development and experimentation expensive thus reducing transparency in distributed inference research. Even when the hardware is available, the lack of standardized benchmarks prevents fair comparison, and exploring hypothetical conditions (e.g., future 6G networks) is impossible without physical access. As a result, researchers lack a unified, controllable, and reproducible way to study how distributed inference algorithms behave across diverse devices and network conditions.

\begin{table*}[!t]
    \centering
    \begin{tabular}{|p{2cm}|p{6cm}|p{3cm}|p{2cm}|p{2.5cm}|}
        \hline 
        \textbf{Framework} & \textbf{Use Case} & \textbf{Network Sim} & \textbf{Integration Complexity} & \textbf{Custom Algorithms} \\ \hline
        DistSim & Analyze training time and device activities in distributed
DNN training & Profiled & N/A & N/A \\ \hline
        SimAI & Simulator for AI large-scale training for analyzing performance data & Advanced & High & Not Supported \\ \hline
        EdgeAISim & Simulating and modelling AI models in edge computing environments & Simple & Medium & Not Supported \\ \hline
        Vidur & LLM inference simulator for deployment conf. & Profiled & Medium & Partial Support \\ \hline
        HERMES & Event-driven, multi-stage inference pipeline simulation and optimization & Simple & Low & Partial Support \\ \hline
        simPy & Discrete-event simulation & (ns.py) & Medium & Supported \\ \hline
        OMNET++ & Discrete-event and network simulation & Advanced & High (C++) & Supported \\ \hline
        sim4DistrDL & DES for distributed training and network sim & Advanced & Medium & Limited Support \\ \hline 
        \textbf{\name} & \textbf{DES for distributed algorithm development, benchmarking and deployment} & \textbf{Profiled} & \textbf{Low} & \textbf{Supported} \\ \hline
    \end{tabular}
    \vspace{4pt}
    \caption{Comparison of existing tools.}
    \label{tab:tools-comparison}
\vspace{-20pt}
\end{table*}

Several discrete-event simulation (DES) frameworks exist that allow users to perform fine-grained network-level modeling and simulation, such as OMNeT++ and ns-3. However, most neural network development and research are conducted in Python, while these tools are primarily implemented in C++, making integration challenging. Even existing Python-based tools such as SimPy lack native support for neural network inference, requiring developers to build custom abstractions and workarounds to adapt existing architectures.

In this paper, we propose \name, short for unified inference—a discrete-event simulation (DES) framework for developing distributed AI models that synchronizes only on network primitives (e.g., send, recv, all-reduce). This design eliminates rollback overhead and enables accurate performance modeling across devices and networks through a unified simulation and deployment API. \name is implemented in Python and supports deployment via PyTorch Distributed, allowing seamless integration with existing models. It treats distributed inference as a first-class citizen, providing abstractions and tools that make benchmarking and profiling more accessible without the need to manually configure complex device arrays or fine-tune network parameters. Moreover, \name ensures that communication is free of race conditions and deadlocks, providing deterministic and reproducible execution across distributed environments. Our contributions are as follows:

\begin{enumerate}
    \item We propose \name, a DES framework for developing distributed AI architectures. \name provides tools that enable developers to simulate, emulate, deploy, and distribute their own developed models across various compute platforms and network conditions.

    \item Leveraging \name, we implement and evaluate the most popular transformer based parallelization schemes, TP, PP, TP-PP Hybrid and Voltage \cite{hu_when_2024}, and demonstrate that our framework can simulate within 98.6\% accuracy, compared to real deployments on varying hardware platforms and network conditions.
    
    \item Through a case study, we show how profiling tools in \name guided the discovery, evaluation, and deployment of \textit{Kilovolts}, an optimization that overlaps computation and communication in the Voltage parallelization scheme, leading to an inference speedup of up to 16\%.
\end{enumerate}
\vspace{-5pt}

\section{Background and Motivation}

\subsection{Distribution of AI Models}

As AI models grow larger, running them on multiple devices has become the de facto standard on high-performance settings \cite{zhong_distserve_nodate}. Recently released models can't fit on a single GPU as they require more than hundreds of GBs of memory. Moreover, many mobile, IoT, and edge scenarios uses hardware with even less compute. Therefore consumers are forced to choose from smaller models and sacrifice performance. There has been various techniques such as Quantization \cite{lin_awq_2024} or Pruning \cite{liu_elastiformer_2024} \cite{liu_flowspec_2025} that try to perform efficiently under limited compute. However they work under the assumption that on-device compute is fixed. We believe it is possible to increase compute by utilizing nearby devices by distributing the model or workload across devices, which would democratize access to more powerful models.

\subsection{Simulation of Distributed Algorithms}

Most common tools for simulating distributed systems and machine learning workflows has been listed in \autoref{tab:tools-comparison}. 



First, simulations of client and server load include Vidur \cite{agrawal_vidur_nodate}, EdgeAISim \cite{nandhakumar_edgeaisim_2024}, SimAI \cite{wang_simai_nodate}, DistSim \cite{lu_distsim_2023} focus on balancing the load coming to the servers, tuning parameters such as batching policy, model parallelism strategy, hardware and model type. One short-coming of such simulators is that they are not running real models underneath but rather predict runtime and behavior based on pre-collected metrics, such as memory usage of different KV-cache sizes and models. Those models come in handy when researchers use them to optimize their compute clusters that serve inference requests to users, however they only support a given preset of models and strategies supported by the simulator. 


Second, performance predictive models \cite{zhang_nn-meter_2021} usually work well in single-device deployments. They utilize kernel-level mappings and optimizations, execution graphs, hardware specs and memory layout to predict time, memory, and power of LLM runtime on varying platforms. However, predicting these metrics in a multi-device setting is difficult due to the difficulty in modeling communication.


Third, pipeline optimizers such as DistSim or HERMES \cite{bambhaniya_understanding_2025}, focus on optimizing the training of AI models by modeling them as pipelines. They focus on batch sizes, GPU utilization, backward passes, filling pipeline bubbles \cite{arfeen_pipefill_nodate}, pre-fill step or RAG retrievals. However they do not provide an easy API for users to develop their custom algorithms. 


\vspace{-5pt}

\subsection{Discrete Event Simulation}

Discrete event simulation, common in network simulators, models the system as a set of states and state transformations based on internal clocks.
Conservative approaches simulate the next time step only when all causalities are resolved and it is safe to do so. In optimistic simulation, individual systems can move their clock forward and roll-back if some interaction happens. But for neural networks, rollbacks have high computation cost.

Currently, DES is used in deep learning frameworks for simulating and optimizing training workloads such as sim4DistrDL \cite{liu_discrete-event-based_2022} or inference workloads such as HERMES. DES has shown to be useful for accurate training simulation without access to real resources. 

Discrete event network simulators, such as OMNET++, can model computer communication costs with great precision, but do not provide an easy way to integrate existing AI models. This is for two reasons. First, most AI models are developed and prototyped in python, while network simulators are often written in C/C++. Second, many of these simulators require developers to extend their interfaces, which is difficult for integrating existing models.

\subsection{Motivation}

Existing tools are proven to be useful in different settings such as single-device runtime prediction or training pipeline optimizations. However, they do not capture the complexity of modeling and evaluating distributed algorithms. Also, existing DES lack an easy way to integrate with AI models. Looking back, existing work on distributed algorithms lack common benchmarks or straightforward ways to replicate their studies.

Also static analysis falls short when predicting runtime of distributed algorithms, most notably due to two factors. \textbf{Batching Efficiency:} Analytical models assume linear processing. They fail to capture the GPU's sub-linear scaling (tiling effects), leading to massive overestimation of latency at high loads. \textbf{Variable Payloads:} Analytical models typically assume fixed network payloads. They fail to capture varying transmission times caused by dynamic tensor shapes. Therefore a \textit{DES} based simulator is needed for accurate prediction and high adaptability.

Our goal is to provide a \textbf{unified} interface and benchmark for development, simulation, evaluation and deployment of distributed AI models that is calibrated and tested against real world deployments.

\section{Design}
 
\subsection{Simulation Engine}

\name runs a discrete-event simulation engine where devices and networks are modeled as logical processes with independently running local clocks. Each process synchronizes either at network events or programmer-defined yield points. \name executes real code instead of simulation, it captures the underlying kernel-level operations and execution graphs as close to the original execution.


\begin{table*}[h]
\centering
\caption{\name's Supported Distributed Communication Operations}
\label{tab:dist_comm_ops}
\begin{tabular}{@{}llll@{}}
\toprule
\textbf{Operation} & \textbf{Hops} & \textbf{Total Bandwidth} & \textbf{Description} \\ 
\midrule
\texttt{broadcast} & $N-1$ & $(N-1) \times M$ & 
\begin{tabular}[t]{@{}l@{}}
One node sends data to all other nodes in the network
\end{tabular} \\[0.5em]
\texttt{all\_gather} & $N-1$ & $N \times (N-1) \times M$ & 
\begin{tabular}[t]{@{}l@{}}
Each node shares its data with all other nodes
\end{tabular} \\[0.5em]
\texttt{all\_reduce} & $2(N-1)$ & $2(N-1) \times M$ & 
\begin{tabular}[t]{@{}l@{}}
Performs reduction operation (sum, max, etc.) across all nodes
\end{tabular} \\[0.5em]
\texttt{send} & $1$ & $M$ & 
\begin{tabular}[t]{@{}l@{}}
Point-to-point transfer from one
node to a specific destination
\end{tabular} \\[0.5em]
\texttt{recv} & $1$ & $M$ & 
\begin{tabular}[t]{@{}l@{}}
Point-to-point receive from a
specific source node
\end{tabular} \\
\bottomrule
\end{tabular}

\vspace{0.5em}
\small
\textit{Note:} $N$ = number of nodes, $M$ = message size. All gather and reduce are ring-based algorithms and use \texttt{send}/\texttt{recv} underneath.
\end{table*}



\begin{figure}[t!]
    \centering
    \includegraphics[width=0.9\linewidth]{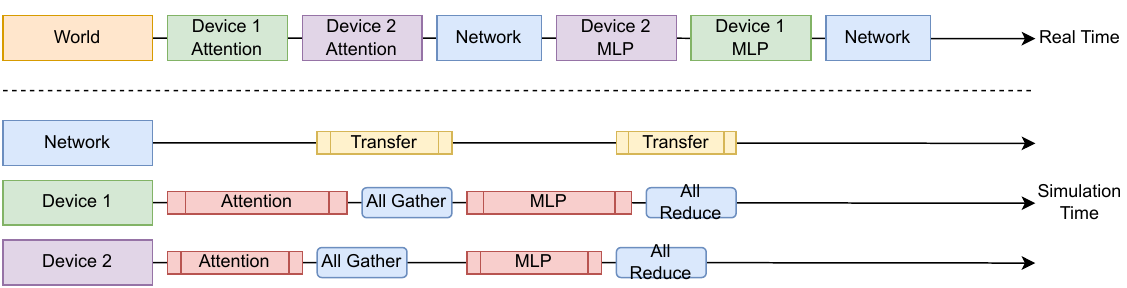}
    \caption{Discrete Event Simulation Overview}
    \label{fig:des}

\vspace{-20pt}
\end{figure}

\noindent
\textbf{Synchronization at interaction:} Interaction events are defined as networked operations, such as sending a message, receiving a message, running all-gather, all-reduce or broadcast operations. Such interactions schedule a network operation at the engine level.
\textbf{Deadlock detection:} The engine runs the simulation while preserving the causal order of events based on network dependencies as shown in \autoref{fig:des}, which detects deadlocks. The causal order also ensures that the simulator doesn't need to rollback events and can accurately simulate the interactions between devices.

\subsection{Simulation of Networked Operations}

The simulation engine models distributed communication operations at a granular level, tracking how bandwidth is dynamically allocated across devices as operations start and finish. Since distributed algorithms depend on network latency and bandwidth, users can either provide these parameters directly or use the built-in profiler to measure them from the actual network. We evaluate the accuracy of both profiling and simulation in \autoref{sec:net-sim}.

\subsection{Simulation Modes and Deployment}

The simulation engine has two modes.

\begin{enumerate}
    \item \textbf{Host Emulation:} The emulation runs on a host device. Each simulated device runs the AI model on a separate logical process (a light-weight thread). 
    


    \item \textbf{Deployment:} The application can be directly deployed to run or benchmark real-life performance on a specific hardware without any code changes. The unified API ensures there are no discrepancy between simulation runs and deployment, ensuring the algorithmic correctness. 
\end{enumerate}

\subsection{Profiling Devices for Simulation}
\label{subsec:profiling}

\name enables profiling device runtime, kernel launches and network events real-time and exporting the simulated timeline using chrome-traces format. \name also integrates directly with the PyTorch profiler to collect low-level traces.

\textbf{Heterogeneity:} To estimate the runtime of a device, users can either specify a slowdown factor or directly run the simulation on the target device type for more accurate results. The slowdown factor allows users to approximate the performance of hardware they do not have access to at scale. Furthermore, users can assign unique slowdown factors to specific kernels or code regions, enabling the emulation of heterogeneous systems with varying scaling characteristics

\section{Results and Discussion}
\label{results}

\name supports a \textit{bring your own architecture} development model. For evaluating our engine's accuracy, we have integrated a text-generation transformer model LLama3 and a multi-modal vision model CLIP-ViT \cite{radford_learning_2021}. 


\subsection{Simulation Overhead}





In this section, we evaluate the impact \name's overhead on its ability to accurately profile inference time. The main source of overhead is thread switching, which occurs in all discrete event simulators when the engine stops and resumes different processes. Next, we describe our setup to evaluate the impacts of thread switching.

\noindent
\textbf{Setup.} Our experiments were run on Jetson Orin Nano and an HPC cluster, 80GB NVIDIA A100 SXM4 with 128GB RAM allocation and 16 vCPUs of Intel(R) Xeon(R) Gold 6338 CPU @ 2.00GHz with the model LLama 3.2-1B and 3B.

For LLMs, the auto-regressive nature of text-generation creates complexities in predicting normalized run-times. First of all, as the generated text is auto-regressive, the generation task gets more computationally expensive over time. Secondly, if the initial prompt is too long, it causes a huge imbalance in computation over time, as kv-caching causes next token generation to be orders of magnitudes faster than the initial pre-fill. Therefore experiments have been designed with different text size and max next token counts to ensure the simulator is tested under different conditions. In order to get stable results and ensure the model doesn't stop generating tokens, the LLM is asked to count numbers without stopping. We have altered the amount of numbers in the prompt.

\begin{table}[b]\centering
\vspace{-20pt}
\caption{Many token prediction experiment regression results}
\begin{tabular}{lcccccc}
\hline
 & \textbf{coef} & \textbf{std err} & \textbf{PI} & \textbf{R\textsuperscript{2}} \\
\hline
\textbf{Const}        & 0.054 & 0.000 & - & 0.779 \\
\textbf{Max tokens} & 12.19  & 0.001 & 24.91 & 0.325 \\
\textbf{Yield count}  & 0.191 & 0.000 & 0.000 & 0.999 \\
\textbf{Device count} & 0.000 & 0.000 & 0.000 & 0.999 \\
\hline
\end{tabular}
\label{table:autoreg}
\end{table}



\noindent
\textbf{Results.} Latin Hypercube Sampling of text-length between 0 and 1000, and thread switch (yield) probability between 0 and 1 was used to get maximum coverage for our auto-regressive performance benchmark. The same experiment was conducted with 1-device, 2-device and 4-device simulations. The effect of device count, number of yields and sequence length on the model runtime was compared using ordinary least squares (OLS) regression. Results in \autoref{table:autoreg} for the autoregressive task show number of yields or simulated devices were statistically significant ($p < 0.01$), however doing permutation importance (PI) and partial $R^2$ analysis shows that their overhead is practically insignificant.

The same experiment was repeated for the prefill task. This time, sequence length varied between 0 and 2000. The same LHS simulation and OLS regression was run again. This time, a quadratic relationship between time and prompt length was found. Parameters yield and device count also did not have practically significant effect on the runtime.

\subsection{Network Simulation Accuracy}
\label{sec:net-sim}

We have deployed \name in a real setting to compare time spent on network to the simulation to ensure accuracy of runtime predictions. Three experiments were conducted on, 8-node HPC cluster (A100 GPU, InfiniBand), 8-node HPC cluster (A100 GPU, 1 Gbps Ethernet), 4× Jetson Orin Nano DevKits (1 Gbps Ethernet).

Infiniband connection used the NCCL backend for synchronizing data between distributed processes, whereas others used Gloo. Each experiment has started on 2 devices and went up to 8 devices (4 for Jetson). On each hardware, two types of experiment was run. The first experiment is designed to accurately model the network parameters, latency and bandwidth. The second experiment was used for evaluating our simulation results in a real inference setting.

\begin{enumerate}[itemsep=0pt, topsep=2pt]
    \item All-gather of a tensor with sizes 1B to 200MB.
    \item Running TP among varying text size and tokens.
\end{enumerate}

For the first experiment, we ran 35 different package sizes with 5 repeats each. The network's perceived latency and bandwidth was found by fitting a simple curve with two parameters using \textit{Ridge Regression}. The latency relationship is modeled as in \autoref{tab:dist_comm_ops}.

\autoref{table:mape-perf} shows that the network latency trends were reproduced with near-perfect accuracy ($R^2 \approx 1$), achieving an average error of ±2\% under high-speed and ±17\% under ultra-high speed networks.

\begin{table}[t!]
\centering
\caption{Model performance comparison (R\textsuperscript{2} and MAPE).}
\begin{tabular}{lccc}
\toprule
\textbf{Hardware} & \textbf{Backend} & \textbf{R\textsuperscript{2}} & \textbf{MAPE} \\
\midrule
HPC & NCCL & 0.9704 & 0.1783 \\
HPC & Gloo & 0.9999 & 0.0248 \\
Jetson Orin & Gloo & 1.0000 & 0.0091 \\
\bottomrule
\end{tabular}
\label{table:mape-perf}
\end{table}

\begin{table}[t!]
\vspace{-10pt}
\caption{Estimation accuracy of real time spent on network}
\centering
\begin{tabular}{lccccc}
\toprule
\textbf{Hardware} & \textbf{Backend} & \textbf{R\textsuperscript{2}} & \textbf{MAPE} \\
\midrule
HPC & NCCL & 0.78-0.90 & 0.2065-0.3626 & \\
HPC & Gloo & 0.9860 & 0.0536 & \\
\bottomrule
\end{tabular}
\label{table:real-life-accuracy}
\vspace{-12pt}
\end{table}

In the second experiment, we executed the TP algorithm, measured the total duration of data transfers and compared these empirical results against the engine's predicted performance. Because production HPC clusters utilize shared network and hardware resources, achieving perfectly consistent results is challenging due to contention and noise. Despite these environmental variables, our simulation accuracy remains high, as summarized in Table~\ref{table:real-life-accuracy}. 



\subsection{Advanced Scenarios}

To show advantage of \textit{DES}, we also evaluated \name with PP and Poisson arrival of requests that are continuously batched. We selected the Poisson distribution as it is the standard proxy for realistic production traffic, where stochastic inter-arrival times create burstiness that forces the system into dynamic batching sizes. Our results show $<$10\% average error on predicting average delay, whereas static analysis fails to scale. In contrast, the analytical baseline (M/D/1) suffered $>$100\% error. Furthermore, \name maintains high accuracy across non-stochastic workloads, yielding ~2\% error for PP and ~5\% error for a hybrid TP+PP parallelization scheme. These results show the engine can accurately simulate tricky synchronization and communication patterns that simple theoretical models miss.
\section{Case Study: Developing a Communication Aware Position-Wise Partitioning Scheme}

\name was built to study improving efficiency of various parallelization schemes. Therefore in order to evaluate efficiency of the simulator, we would like to share our optimization of the \textit{Voltage} algorithm~\cite{hu_when_2024}, which we call \textit{Kilovolts}, using computation communication overlap and how we validated it using our simulator.

Voltage is a distributed inference system designed for edge devices that reduces bandwidth consumption by up to 4× compared to conventional tensor parallelism. It achieves this reduction through a position-wise layer partitioning strategy. In each transformer block, the output of the previous layer $x$ is distributed across all devices. Each device computes its own partition of the transformer layer and then performs an all-gather operation. The aggregated result corresponds to the output of current transformer block, which is subsequently passed as input to the next layer.

\begin{figure}[b]
    \vspace{-20pt}
    \centering
    \includegraphics[width=0.8\linewidth]{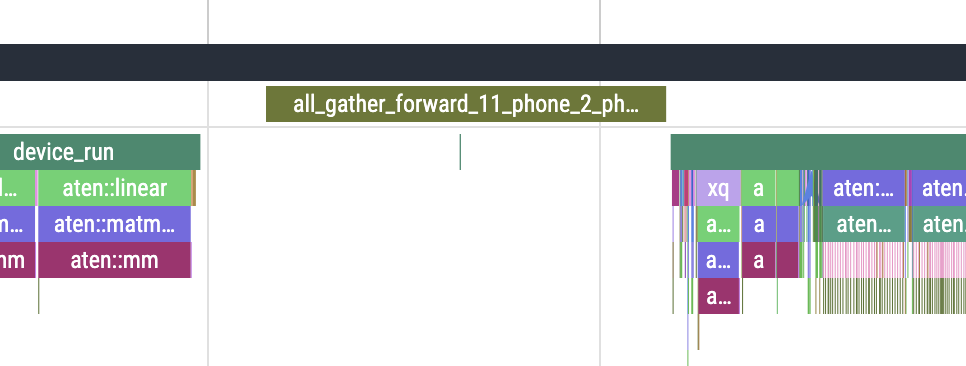}
    \includegraphics[width=0.8\linewidth]{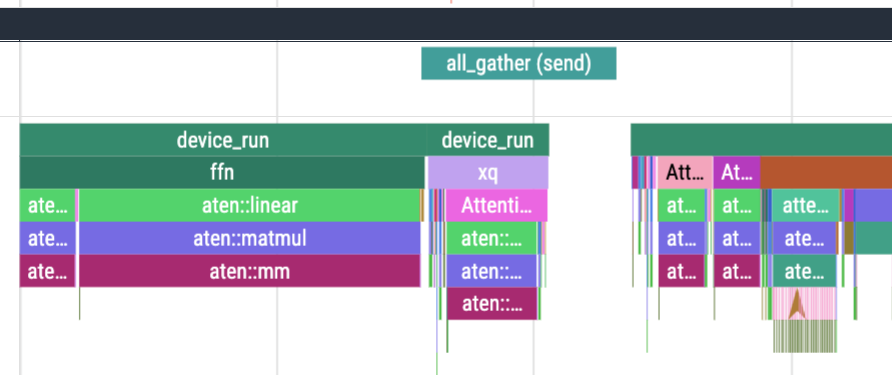}
    \caption{\name Trace Tool showing Voltage~\cite{hu_when_2024} (top) and the improved \textit{Kilovolts} algorithm (bottom) with the communication overlap on \textit{xq}.}
    \label{fig:placeholder}
    \vspace{-10pt}
\end{figure}

\noindent
\textbf{Case Study.} Our simulator didn't require any major changes on the neural networks unlike the other python simulators that depend on generators. Therefore LLama3 was directly forked and integrated into the simulator with a thin wrapper. The Voltage algorithm implements the attention as follows:

\vspace{-10pt}

$$
    A_p(x) = \text{softmax} \left( \frac{(x_pW_Q)(xW_K)^T}{\sqrt{F_H}} \right) (xW_V )
$$

 Using the \name trace tool (Figure~\ref{fig:placeholder}), we saw that the computation of $x_pW_Q$ runs after the all gather. However calculating $x_pW_Q$ doesn't need to wait for the all-gather, as $x_p$ is already available from the previous layer without the all gather operation. All gather is only required for collecting remaining partitions to fully assemble $x$. The time gained from this overlap can be calculated as $max(0, T_{transfer} - T_{xW_q})$. Next, we study how transfer and calculation speed changes between different input shapes and network conditions using our simulator, which showed a potential speed-up between 1\% to 10\%. 


For comparison, we deployed Voltage and Kilovolts on four Jetson Orins and varied the network bandwidth and latency. We observe speedups of up to 5\% with CPU (\autoref{tab:voltage_cpu_results}) and up to 16.1\% with GPU execution (\autoref{tab:voltage_cuda_results}), which matches closely with simulations. Deep-profiling revealed that fluctuations in measurements are caused by internal queueing and buffering of asynchronous collective operations.

\begin{table}[h]
\centering
\vspace{-12pt}
\caption{Speed up of \textit{Kilovolts} on Orin CPU}
\vspace{-5pt}
\label{tab:voltage_cpu_results}
\small 
\setlength{\tabcolsep}{4pt} 
\begin{tabular}{ccccccc}
\toprule \textbf{Input Len} & \textbf{Bw.} & \textbf{Lat.} & \textbf{Time} & \textbf{Speedup} \\
\textbf{} & \textbf{(Mbps)} & \textbf{(ms)} & \textbf{(s)} & \textbf{(\%)} & \\
\midrule
8    & 100  & 10  & 1.729 & $[-8.07\%, 2.13\%]$ \\
32	 & 1000 & 1   & 1.563 & $[-11.81\%, 4.94\%]$ \\
256  & 100  & 10  & 12.22 & $[1.24\%, 3.48\%]$ \\
\bottomrule
\end{tabular}
\vspace{-5pt}
\end{table}

\noindent
\textbf{Deployment Experience.} Our deployment experience across varying network conditions highlights the need for a simulator, like \name, that allows developers and researchers to easily evaluate system performance under diverse network conditions.

While varying network parameters, we found that the Linux kernel provided with the Jetson boards did not include the \textit{htb}, \textit{tbf}, or \textit{netem} modules. As a result, developers were required to rebuild the kernel and reflash all four devices to enable traffic control. Using middleware solutions such as toxiproxy was also infeasible, since the Gloo backend relies on ephemeral ports and does not allow specifying fixed target ports for communication—an open issue in the Gloo repository that has remained unresolved for over seven years. We also experimented with redirecting a port range to toxiproxy using NAT table rules, but this approach was incompatible with Gloo’s multiplexed connections. The only viable solution to control network bandwidth was to employ an OpenWRT router and use \textit{netem} and \textit{tbf} for traffic shaping. 

\begin{table}[h!]
\centering
\vspace{-10pt}
\caption{Speed up of Orin on CUDA}
\vspace{-5pt}
\label{tab:voltage_cuda_results}
\small 
\setlength{\tabcolsep}{4pt} 
\begin{tabular}{ccccccc}
\toprule \textbf{Input Len} & \textbf{Bw.} & \textbf{Lat.} & \textbf{Time} & \textbf{Speedup} \\
\textbf{} & \textbf{(Mbps)} & \textbf{(ms)} & \textbf{(s)} & \textbf{(\%)} & \\
\midrule
269  & 100  & 1  & 1.253 & $[-3.11\%, 3.15\%]$ \\
269  & 100  & 10 & 1.710 & $[-1.24\%, 4.91\%]$ \\
269  & 1000 & 1  & 0.428 & $[-7.43\%, 4.34\%]$ \\
1002 & 100  & 1  & 3.890 & $[-5.60\%, 4.99\%]$ \\
1002 & 100  & 10 & 4.015 & $[-2.66\%, 7.59\%]$ \\
1002 & 1000 & 1  & 1.090 & $[-19.1\%, 16.1\%]$ \\
\bottomrule
\end{tabular}
\vspace{-10pt}
\end{table}

\section{Limitations and Future Work}


The simulator was originally developed to support our research on studying, replicating, and discovering new distributed inference algorithms. Consequently, several engineering decisions were made to suit our specific needs, such as focusing primarily on transformer-based models and implementing a custom network model rather than adopting a sophisticated simulator like ns-3. Nonetheless, our framework supports the integration of diverse model architectures and network backends within the DES environment. We hope this flexibility will enable researchers to bring their own models and tools according to their experimental requirements.

Currently, the tool runs on the host device to simulate distributed execution. If the host device has limited memory, it may not be feasible to run the full model locally. In such cases, users can either share weights across simulated devices or execute the simulation on higher-end hardware while applying a slowdown factor to approximate the performance of the target device. However a static slowdown factor might not capture performance characteristics of different devices. We are also developing a remote target mode, in which the DES engine can attach to external devices over the network. Because the engine makes no assumptions about the underlying hardware, this mode will enable real-time performance characterization on actual remote devices, eliminating the need to run lightweight threads on the host. 


\vspace{-5pt}

\section{Conclusion}
We presented \name, a discrete event simulator for developing distributed AI algorithms. While the current framework provides a robust foundation, future research should focus on improving low-level runtime approximations across heterogeneous devices and overcoming host memory constraints. In addition, analyzing behavior under high-speed network configurations could lead to more accurate models for HPC environments. Ultimately, \name can serve as a reliable and scalable platform for developing, deploying and benchmarking distributed algorithms and for reproducing both prior and emerging studies, helping bridge the gap between theoretical and practical performance limits.

\bibliographystyle{IEEEtran}
\bibliography{references.bib}

\end{document}